# The SIDE dual VIS-NIR fiber fed spectrograph for the 10.4 m Gran Telescopio Canarias

O. Rabaza*[a], H.W. Epps[b], M. Ubierna[a], J. Sánchez[a], M. Azzaro[a], F. Prada[a]
[a]Institute of Astrophysics of Andalucia (CSIC)
[b]UCO/Lick Observatory UC Santa Cruz

**ABSTRACT**

SIDE (Super Ifu Deployable Experiment) is proposed as second-generation, common-user instrument for the GTC. It will be a low and intermediate resolution fiber fed spectrograph, highly efficient in multi-object and 3D spectroscopy. The low resolution part (R = 1500, 4000) is called Dual VIS-NIR because it will observe in the VIS and NIR bands (0.4 ~V 1.7 microns) simultaneously. Because of the large number of fibers, a set of ~10 identical spectrographs is needed, each with a mirror collimator, a dichroic and two refractive cameras. The cameras are optimized for 0.4 - 0.95 microns (VIS) and 0.95 - 1.7 microns (NIR) respectively.

## 1. INTRODUCTION

Within the context of the SIDE instrument, the spectrograph that will provide moderate resolutions 1500 and 4000 has been designated as the Dual VIS-NIR spectrograph. In this paper we present preliminary optical designs, for the collimator, beamsplitter, dispersive elements, order-sorting filters, the VISIBLE camera optics and the NIR camera optics for such spectrograph. We provide quantitative image analyses and polychromatic spot diagram evaluations for the system.

We assume each fiber bundle containing 7 closely packed 0.5-arcsec fibers, illuminated at the telescope focus by microlenslet arrays such that their output into the collimator can be specified to f/5.0. The 7 fibers from each unit will be placed side-by-side, as close as possible one to another, to form a pseudo-slit (for each object) that is 0.5-arcsec wide along the dispersion direction, considering a fibre core of 120 micron and a buffer diameter of 180 micron.

A 4096 x 4096 with 15 microns per pixel LBNL CCD and a 2048 x 2048 with 18 microns per pixel HAWAII 2 RG or equivalent array has been considered as detectors. The required R = 1500 and 4000 resolution could be attained with a collimated beam diameter of about 100 mm. A collimator field radius of 6.0 degrees was adopted such that the total number of objects in MOS mode is about ~ 100 per spectrograph. The pixel sampling along the dispersion and spatial direction becomes 4.72 pixels/pseudo-slit-width for the VIS arm and 2.36 pixels/pseudo-slit-width for the NIR arm.

These parameters, led to a required NIR camera focal length of about 175.4 mm and a VIS camera focal length of about 292.3 mm both cameras field radius of about 8.5 degrees, which covers the corners of the array. Moreover, in both arms we adopted a folded optical path to reduce the price of the gratings and increase the total transmission, because with the linear optical path solution the VPH gratings would be thicker.

*ovidio@iaa.es; phone (+34) 958 230 623; fax (+34) 958 814 530; iaa.es

Table 1. Science requirements.

| SCIENCE REQUIREMENTS | |
|---|---|
| **VIS** | |
| Resolutions | 1500, 4000 |
| Wavelength Range | 0.40-0.95 µm |
| **NIR** | |
| Resolutions | 1500, 4000 |
| Wavelength Range | 0.95-1.7 µm |
| FOVs | |
| MOS at Nasmyth | 20' Diameter |
| SIFU at F-Cass | 28.5"x 28.5" |
| mIFU at F-Cass | 10' Diameter |

## 2. SPECTROGRAPH OVERVIEW

### 2.1 Quantitative Basis for the Spectrograph Parameter Selection

The fundamental parameters that define the characteristics of the SIDE Dual VIS-NIR spectrograph are summarized in Table 2.

Table. 2. Goal parameters that define the characteristics of SIDE Dual VIS-NIR spectrograph.

| | |
|---|---|
| Telescope Diameter | 10000 mm |
| Aperture per single fiber | 0.5 arcsec |
| Beam Diameter | 100 mm |
| Collimator Field Radius | 6 degrees |
| Resolutions | 1500, 4000 |
| Central Wavelengths | 0.56, 0.74, 0.45, 0.56, 0.69, 0.85, 1.06, 1.27, 1.32, 1.58 |

Since it was clear that the fiber inputs to the collimators would be aligned along the pseudo-slits, perpendicular to dispersion, and some space would be required between the fibers for cladding, etc., we chose to assign one "dark space" of equal width to each fiber so as to form imaginary "units" along each of the pseudo-slits. In practice, the number of fibers that can be accommodated along each pseudo-slit will depend on the details of their construction, cladding, and the amount of dead space provided between object spectra. Those details must be established in a subsequent study but the number of units are the total amount of pseudo-slit length available for the purpose in a convenient way. Our initial calculations led quickly to two results, which were fundamental to all of the numerical work that followed:

The telescope imaging scale (determined by its focal length) can be modified prior to the collimators with no appreciable impact on the optics that follow, provided the aberrations introduced by doing so are negligible. For example, the 20.0-arcmin field corrector could modify the scale. It could also be modified by using individual microlenses or microlenslet arrays as inputs to the fibers that couple the telescope f.o.v. to the collimators. We assumed that such fore-optics would in fact be used, such that we could safely define a new scale without designing the fore-optics ahead of time. As an initial approximation we chose a scale of 242.65 microns/arcsec, which provides an f/5 input to the fibers.

It did not appear possible to design spectrographs for SIDE that could accept the stipulated 1.5-arcsec input fiber size without adopting collimated beam diameters that would be prohibitive in cost and instrument volume. Thus, some form of image slicing would be required. We chose to proceed on the assumption that the fiber input for each astronomical object in the f.o.v. would be a "hex-pack" (see Figure 1) containing one central 0.5-arcsec diameter fiber, surrounded by 6 additional 0.5-arcsec diameter fibers to form a symmetric hexagonal shape. We assumed that these hex-packs would be custom-designed and manufactured for maximum packing factor, such that the equivalent circular diameter would be about 1.5 arcsec, with each 0.5-arcsec fiber core being about 120.0 microns in diameter.

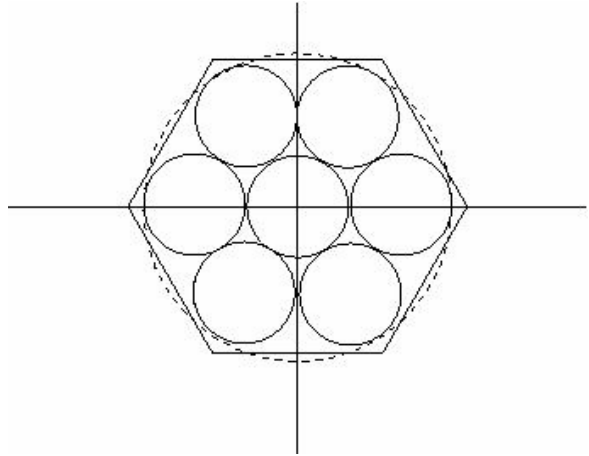

Fig. 1. Hex-pack unit configuration for MOS mode.

An interesting consequence of the 1st point listed above is the fact that the sizes, geometries and designs for the dispersing elements and the cameras do not depend at all on the *choice of scale* at the fiber input end. Thus, if that choice should need to be modified after those designs are completed, no work will have been lost and no harm will have been done.

**2.2 Parameters Selected for the SIDE Dual NIR arm.**

A (2048 x 2048 by 20-micron) Raytheon array was initially specified as the detector. It was assumed that dispersion would be provided with grisms or VPH gratings. If so, the central anamorphic factor would be 1.0 with a range of about +/- 0.05 at the red/blue ends of the spectra. A collimated beam diameter of 100.0 mm was adopted as a compromise that would be affordable to build, would provide adequate resolution and would not dilute the pixel sampling at the detector to an unacceptably low level. A collimator field radius of 6.0 degrees was selected as a compromise that would provide an adequate total pseudo-slit length but would not cause the camera's required field radius to become excessive and would also not dilute the pixel sampling at the detector unacceptably.

The pixel sampling is 4.72 pixels/arcsec (2.36 pixels/pseudo-slit-width) in the dispersion direction, which is quite coarse but may just barely be acceptable.

These parameters, led to a required camera focal length of 194.9 mm and a field radius of 8.5 degrees, which covers the corners of the array. The dispersion values required to accommodate central wavelengths in the (1.3 to 1.5)-micron range, with resolution in the R= (1500 to 4000) range, appear to be attainable with practical VPHs. The selection criteria will include transparency, and especially *availability* in the large sizes required. An attempt will be made to minimize the prisms angles to the extent possible.

The VPH entry face(s) will have to be roughly 112 mm wide, such that their depth(s) along the optical axis could reach roughly 150 mm. If the nominal collimator exit pupil is placed at the mid-point, the camera's entrance aperture vertex would have to be located at least 110 mm beyond that plane. Conservatively, We adopted an entrance pupil distance of 130.0 mm for the preliminary NIR camera design. That allows some extra room for mounting hardware, and it led to a predesign estimate for the camera's entrance aperture diameter of about 146 mm (f/1.33, underfilled).

As we will see below will be possible to reduce the entrance pupil distance would lead to cameras that were smaller, optically slower and somewhat easier to design and build.

An alternative (2048 x 2048 by 18-micron) Hawaii-2RG array may be a more realistic choice for the detector. With all of the other parameter choices listed above remaining the same, led to a required camera focal length of 175.4 mm. Since all of optics ahead of the camera must remain the same, the aforementioned 146-mm entrance aperture diameter results in an even faster, harder to design (f/1.20, underfilled) NIR camera.

### 2.3 Parameters Selected for the SIDE Dual VIS arm.

A (4096 x 4096 by 15-micron) LBNL CCD or equivalent array has been specified as the detector. The 100.0-mm collimated beam diameter, VPH gratings with a central anamorphic factor of 1.0 and a 6.0-degree collimator field radius were assumed for the VIS optics for the same reasons mentioned above for the NIR optics. The resulting pixel sampling of 4.72 pixels/pseudo-slit-width in the dispersion direction is generous. Note that since the same collimator will be used for both the NIR and VIS spectra *simultaneously*, with the beams being split by a dichroic mirror placed ahead of the dispersing elements, it is not practical to increase the VIS collimator field radius. These parameters led to a required camera focal length of 292.3 mm and a field radius of 8.5 degrees, which covers the corners of the array. The dispersion values required to accommodate central wavelengths in the (0.5 to 0.85) micron range, with resolution in the R= (1,500 to 4000) range, appear to be attainable with practical VPHs. Thus, the conclusions about entrance pupil distance for the VIS camera are the same as for the NIR camera. The estimated 146-mm entrance pupil diameter results in a somewhat slower (f/2.00, underfilled) VIS camera.

## 3. SPECTROGRAPH OPTICAL DESIGN

In the following sections we will show each optical element of the SIDE Dual VIS-NIR Spectrograph taking into account the initials parameters established above.

### 3.1 Slithead

Light enters the spectrograph through fibers, which terminate at the slithead. For the purposes of this study, we have assumed 120 μm diameter fibers with an f/5 output cone. The fibers are stacked vertically to form a long slit and are placed on a radius whose center of curvature coincides with that of the collimator. Additionally, the fibers are aimed in a fanlike pattern outward from the center curvature toward the collimator, so that the central (gut) ray from each fibers strikes the collimator normal to the surface. Thus, the slithead is at the focus of a one-dimensional Schmidt collimator. The SIDE Dual VIS-NIR slit will be 105 mm in length.

### 3.2 Collimator

The collimators for SIDE will be quite simple and effective. The SIDE Dual VIS-NIR collimator is assumed to work at f/5 such that it requires a 500.0-mm focal length to produce the adopted 100.0-mm beam diameter. The only optical element needed is a spherical mirror whose radius is twice the focal length, or 1000.0 mm. The fiber "ends," where light exits, must be placed along a circular arc whose radius is 500.0-mm, with a center point that is also located at the mirror's center of curvature. This geometry will produce a 100.0-mm diameter exit pupil, on a plane perpendicular to the optical axis that passes through the mirror's center of curvature. Thus, in effect, we can think of the collimator as a "correctorless" Schmidt camera working in reverse, whose f.o.v. is confined the plane of the fiber fan (Slithead).

By symmetry, it is apparent that the only aberration present will be spherical aberration, which is independent of field angle. The only optical requirement is that the spherical aberration produces a "spot size" that is small and negligible compared with the diameter of each individual fiber. A Zemax model along these lines was formulated for the SIDE Dual VIS-NIR collimator so as to calculate the aberration spot size by direct ray tracing. The model was illuminated in parallel light, through its "exit" pupil, like a normal Schmidt camera. The resulting spots (on the curved slithead surface) are shown in Figure 2 for several field angles (on axis, 1, 2, 3, 4, 5 and 6) of the full 6.0-degree field radius. The total aberration blur spot is very small as compared with the 120.0-micron diameter fiber shown as a circumscribed circle.

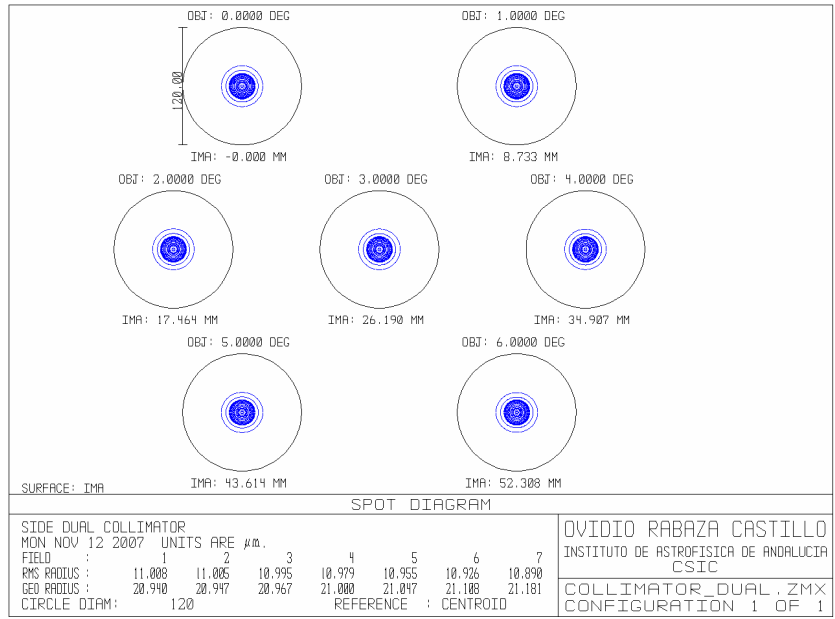

Fig. 2. Qualitative measured of the collimator spherical aberration for several field angles.

The mirror itself is fabricated from a rectangular blank, 100 mm wide, 310 mm tall, and 60 mm thick with a final radius of 1000 mm. The collimator is the largest optic in the spectrograph and drives the overall height of the optical bench.

The collimator forms a pupil at the center of curvature of the mirror, and this is where the gratings are located in each channel in order to minimize their required size.

### 3.3 Beamsplitter

A dichroic beamsplitter divides the incident collimated beam, reflecting the VIS wavelengths portion of the bandpass ($\lambda < 950$ nm) and transmitting NIR wavelengths ($\lambda > 950$ nm). It is fabricated from fused silica, 160 x 160 x 35 mm, with the dichroic coating applied to the incident surface. The coating will reflect the VIS light very efficiently (R > 99%) and transmit the red light somewhat less efficiently (T > 92% average, including the reflection loss at the exit surface, which has a high performance broadband antireflection coating). The 10%–90% zone at the crossover wavelength is approximately 50 nm wide.

### 3.4 Order-Sorting Filters

In the VIS arm we will need only two order sorting filters; Schott GG395 and GG495 whereas in the NIR arm we don't need order sorting filter because the own beamsplitter cuts the superior diffraction orders.

### 3.5 Gratings

As a main conclusion, the proposed VPH option is conceptually a feasible option that should perform as required regarding the resolution. No manufacture difficulties are expected with the current parameters. Minor recommendations are given in order to close the concept and for the following phases.

The current SIDE concept contains two different cameras and dispersive elements. The F/5 collimator supply a 100 mm beam that is splitted at the dichroic. We can see in Figure 3 and Figure 4 an optical layout of the SIDE Dual VIS-NIR working in resolution 1500 and resolution 4000. For change of resolution we will use two grating types, "VPH+Prism" and "Regular VPH". In principle to cover all the spectral range from 0.4 μm to 1.7 μm we will use a total of 10 disperser elements (3 "VPH+Prisms" and 7 "Regular VPHs"), although, depending of the spectral ranges of interest maybe the number of disperser elements will be less.

Table. 3. Main parameters for the VIS dispersive elements.

| Type | Material | Grooves |
|---|---|---|
| VPH+Prisms | S-LAL7 | 550 lines/mm |
| VPH+Prisms | S-LAL7 | 400 lines/mm |
| Regular VPH | N-BK7 | 1950 lines/mm |
| Regular VPH | N-BK7 | 1570 lines/mm |
| Regular VPH | N-BK7 | 1280 lines/mm |
| Regular VPH | N-BK7 | 1030 lines/mm |

Table. 4. Main parameters for the NIR dispersive elements.

| Type | Material | Grooves |
|---|---|---|
| VPH+Prisms | S-LAL7 | 220 lines/mm |
| Regular VPH | N-BK7 | 820 lines/mm |
| Regular VPH | N-BK7 | 690 lines/mm |

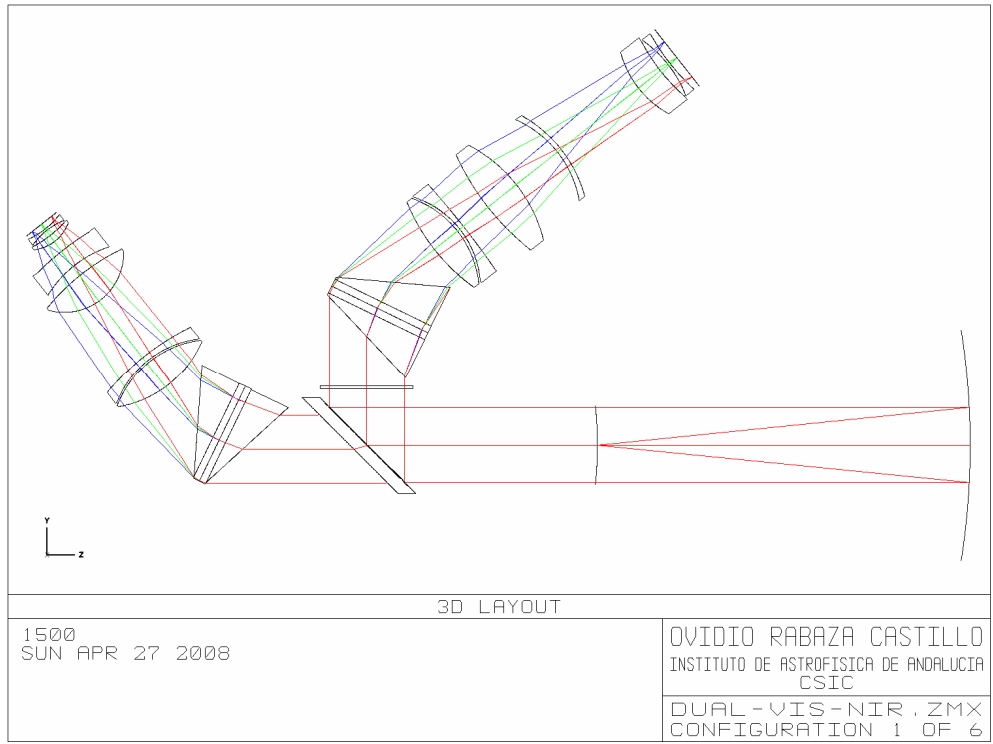

Fig. 3. Optical layout of SIDE Dual VIS-NIR Spectrograph at resolution 1500.

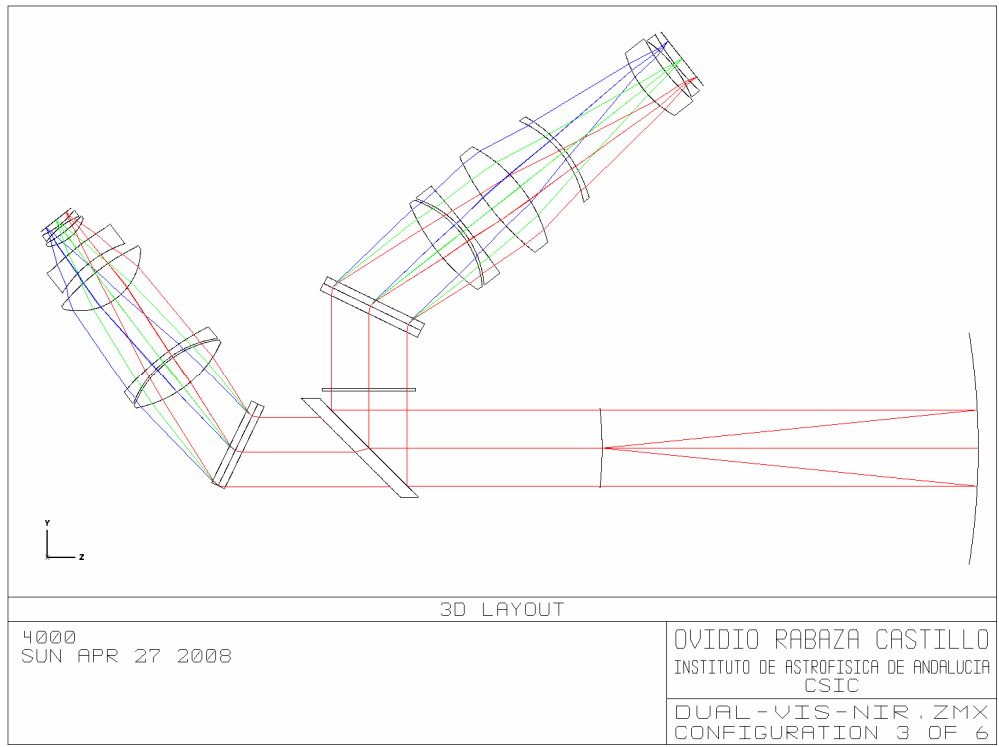

Fig. 4. Optical layout of SIDE Dual VIS-NIR Spectrograph at resolution 4000.

### 3.6 Cameras

**Selection of Infrared Optical Materials and their Characteristic Properties:**

The design of high-performance optics for the traditional (1.0 to 2.4)-micron chromatic range is strongly curtailed by the limited selection of suitable optical materials [1]. However limiting the long-wavelength cut-off for the design to 1.70 microns makes it practical to consider a number of standard optical glasses, which would otherwise have been excluded due to strong absorption beyond 1.8 microns.

The Ohara catalogue was reviewed in order to select available optical glasses which exhibit an internal transmission of 0.994 or greater at a wavelength of 1.40 microns, for a 10.0-mm thickness.

Traditional infrared materials from the aforementioned SPIE reference are included, such as BaF2, CaF2, LiF, Infrared Fused Quartz (called FQTZ), etc. It can be seen that the uniquely small dispersive power of Barium Fluoride (BaF2) makes it extremely attractive as the provider of substantial positive power in a fast, high-performance camera. However its well known hygroscopic tendency (some 100 times worse than CaF2) and its extreme softness (which makes it difficult to polish and to coat) conspire to make it a poor choice for use in SIDE.

For convenience of intercomparison, the data are all given at "room temperature" (T= +25.0 C for the optical glasses; T= +20.0 C for the traditional infrared materials). The preliminary optical designs reported here were also done at room temperature as the detailed temperature distribution through the camera(s) has not yet been specified. Generic refractive indices were used for the various optical materials. The corrections to proper temperature(s) will all be at the level of typical "melt-sheet" corrections. They can be done when the selected designs are updated prior to construction.

One notes also that the camera designs presented make no distinction between "air" and "vacuum." This approximation is safe as refractive index corrections at that small level can also be made at construction design time or even ignored entirely as they always result only in a very small wavelength-independent focus shift, even in optical systems which are much faster optically than the cameras needed for SIDE.

**A 6-Element All-Spherical f/1.20 NIR Camera Lens Design for a 18.0-micron Detector:**

The 175.4-mm focal length specification required for this camera, leads to an even more difficult (f/1.20, underfilled). We started the design from the f/1.33 model and quickly determined that the axial color correction, required to maintain parfocality with wavelength, represented the pivotal issue in the design. We repeated some of our glass exchange experiments with the moderate-to-high dispersion optical glasses and tried to redistribute the positive and negative optical power among the individual lens elements but he found very little that was helpful along those lines. Basically, the faster camera simply causes all of the various curvature and thickness issues to become more extreme. The residual axial color (as a fraction of focal length) remains roughly the same as in the prior (f/1.33) model but the faster (f/1.20) f/ratio in this model causes an increase in the rms image diameters. The smaller 18.0-micron pixels make matters even worse.

We experimented with the addition of another lens element and also with the use of an aspheric surface. These modifications tended to produce designs with somewhat less severe geometry and slightly better image quality but the improvements did not appear to justify the added complexity (and anticipated large construction cost differential). Thus, we did not pursue them further.

It proved possible to retain the relative geometry of the 4th, 5th and 6th lens elements such that the (very thick) 4th lens element is still a good candidate to be used as the vacuum window. As with the f/1.33 camera for the 20.0-micron detector, doing so will provide maximum isolation of the sensitive NIR detector, enabling it to be cooled and well shielded from unwanted background radiation.

When illuminated in perfectly parallel light from the entrance pupil, as described above, this 6-element all-spherical camera shows residual aberrations with an rms image diameter of 19.1 +/- 3.2 microns (1.06 +/- 0.18 pixels) averaged over field angles and wavelengths within the (0.95 to 1.73)-micron passband *without refocus*. 3rd-order barrel distortion is some 0.55% at the edge of the full field (the corners of the detector). Thus, the image quality is quite good.

A scaled drawing of this camera is shown in Figure 5 parallel light rays, represented by the chief ray and the marginal rays for an 8.5-degree field angle, radiate from the entrance pupil on the left, moving toward the right. The rays pass through the lens elements and converge to focus at the flat detector array on the right. The total length from the entrance pupil to the detector is 431.45 mm. The length of the camera itself, from the first lens vertex to the detector is 301.45 mm, which is substantially more than the camera's 175.4-mm focal length. This difference is typical in camera designs of this type and it cannot easily be avoided.

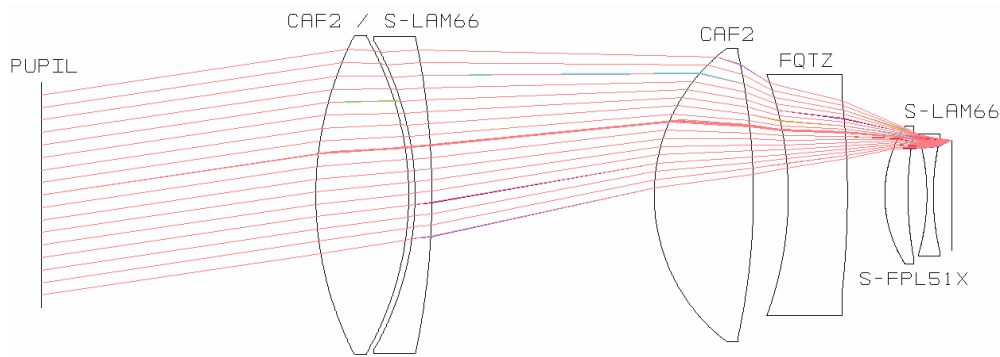

Fig. 5. SIDE NIR Camera optical design for a 2k×2k @ 18 μm detector.

The lens apertures shown are minimum clear apertures. In practice, the lens elements would be made larger in diameter so as to allow sufficient material for mounting purposes, for bevelling and for convenience of optical figuring and coating.

The entrance pupil distance needed taking into account the gratings design will be as maximum 60 mm instead of the 130 mm designed, then the camera's first lens element could become as large as 130 mm in diameter and the final camera's optical speed only will be as fast as f/1.35 (underfilled).

**A 6-Element All-Spherical f/2.0 VIS Camera Lens Design for a 15.0-micron Detector:**

The 292.3-mm focal length specification required for this camera, together with all of the other parameters given for the NIR camera, leads to slower (f/2.00, underfilled) geometry. However the extreme (0.39 to 0.95)-micron spectral range makes the axial color correction even harder to achieve than in either of the NIR camera. The choice of materials is key and that choice is made all the more complicated by reason of the larger number of suitable materials available in this spectral range.

We used the 194.9-mm NIR camera as the starting point and retained CaF2 as the obvious best choice to produce the large positive power provided by lens #1 and lens #3. We left the materials for lens #5 and lens #6 as they were, since their function is primarily to flatten the field, by providing large curvatures near the focal plane where their effects on the axial color are not extreme. Thus, the initial design effort was to assess the color correction interplay between the two doublets, by manipulating the choices of materials for lens #2 and lens #4. This is a somewhat "tricky" process as the interplay between two different dispersion curves can easily lead to false conclusions regarding the "best" glass combination. Trial and error is not a viable approach as there are simply too many possible combinations to investigate.

It became obvious that the color correction needed to be distributed more evenly than had been the case in the NIR designs. For that purpose, we chose lower-index, lower-dispersion glasses for lens #2 and we found that Ohara S-LAL7 appeared to be good compromise, by reason of its excellent transmission in the near-uv as compared with other glasses with comparable indices and dispersive properties.

As the 2nd doublet assumed more of the color correcting responsibility, it became clear that the interaction improved steadily as the group was allowed to evolve into a free-standing quartet. Doing so enabled lens #4 to become more effective by getting thinner, while at the same time, lens #5 improved the system by becoming thicker. In this approach, lens #5 becomes identified as the best candidate to be used as the vacuum window.

Additional glass exchange calculations identified Ohara @PBL26Y as the best choice for lens #4. This is a special "i-line" glass with enhanced UV transmission all the way down to the i-line at 0.365 microns. Additional calculations showed that S-FPL51Y and @PBM2Y made an excellent combination for lens #5 and lens #6. These are also Ohara "i-line" glasses with excellent UV transmission.

We found that by adding a 7th lens element, we could improve the image quality somewhat but the improvement did not appear to justify the added complexity (and anticipated large construction cost differential). Thus, we did not pursue it further. We didn't try an aspheric surface in the VIS camera because we thought that doing so would not improve the image quality enough to warrant the extra cost. Axial color, not severe geometry, is the main residual aberration in the VIS camera and that problem is rarely helped by adding an aspheric surface.

When illuminated in perfectly parallel light from the entrance pupil, as described above, this 6-element all-spherical camera shows residual aberrations with an rms image diameter of 14.5 +/- 4.2 microns (0.97 +/- 0.28 pixels) averaged over field angles and wavelengths within the (0.39 to 0.95)-micron passband *without refocus*. 3rd-order barrel distortion is some 0.13% at the edge of the full field (the corners of the detector). Thus, the image quality is comparable (in pixels) to that in the 175.4-mm NIR camera although somewhat more variable. It appears to be acceptable.

A scaled drawing of this camera is shown in Figure 6 parallel light rays, represented by the chief ray and the marginal rays for an 8.5-degree field angle, radiate from the entrance pupil on the left, moving toward the right. The rays pass through the lens elements and converge to focus at the flat detector array on the right. The total length from the entrance pupil to the detector is 539.35 mm. The length of the camera itself, from the first lens vertex to the detector is 409.35 mm, which is substantially more than the camera's 292.3-mm focal length. This difference is typical in camera designs of this type and it cannot easily be avoided.

The lens apertures shown are minimum clear apertures. In practice, the lens elements would be made larger in diameter so as to allow sufficient material for mounting purposes, for bevelling and for convenience of optical figuring and coating.

The entrance pupil distance needed taking into account the gratings design will be as maximum 60 mm instead of the 130 mm designed, then the camera's first lens element could become as large as 127.3 mm in diameter and the final camera's optical speed only will be as fast as f/2.3 (underfilled).

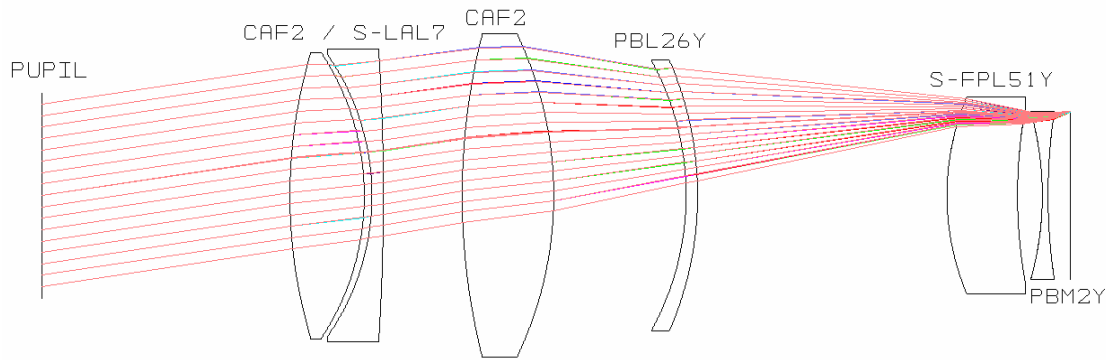

Fig. 6. SIDE VIS Camera optical design for a 4k×4k @ 15 μm detector.

**Detail Optical Performance of the Preliminary NIR and VIS Camera designs:**

It is commonly agreed that if images are more-or-less "round", without extreme bumps, wiggles or tails, and they are more-or-less "smooth", without rings and voids, then they will be pixel-sampling limited if about 80% or more of the energy is included within a (2-pixel x 2-pixel) Nyquist sampling box and 95% or more of the energy is included within a (3-pixel x 3-pixel) sampling box (both centered on the image centroid). If the images were Gaussian in shape, 80% of the energy would be contained within 1.27 rms image diameters and 95% would be contained within 1.72 rms image diameters. The corresponding criteria for quasi Gaussian images to be pixel-sampling limited is that *(roughly speaking)*, they should have rms diameters in the (1.6 to 1.7) pixel range (or smaller). This criterion is no substitute for detailed modulation transfer calculations but it is probably more than adequate for the purpose of examining preliminary optical designs.

A (100 by 107)-mm entrance pupil at an entrance-pupil distance of 130.0 mm was used for the calculations and the cameras were illuminated in perfectly parallel light. It should be noticed that this illumination pattern actually "overdrives" the cameras in the sense that the bluer wavelengths will have smaller anamorphic factors. The distribution of rms image diameters for the optically faster 175.4-mm focal length (18-micron) camera are somewhat larger and more variable but the worst-case rms image diameter is only 1.31 pixels (at the extreme blue wavelength, in the very corner of the detector). Thus, it appears fair to conclude that both of these preliminary NIR camera designs are strongly pixel-sampling limited. Their imaging characteristics appear to be excellent for the intended purpose.

The illumination pattern was the same as described above for the NIR camera and the same comment about "overdriving" the camera applies here as well. It can be seen that the worst case images for the 292.3-mm focal length (15-micron) VIS camera have rms diameters that are somewhat larger than those in the NIR camera, mostly due to the longer focal length and the smaller pixel size. However the worst-case rms image diameter is only 1.60 pixels (at the extreme blue wavelength, in the very corner of the detector). Thus, it appears fair to conclude that the VIS camera design is also strongly pixel-sampling limited. Its imaging characteristics appear to be excellent for the intended purpose.

The expected optical performance of these cameras is further illustrated with *polychromatic* spot diagrams, in Figure 7 for the 18-micron NIR camera and Figure 8 for the VIS camera. All of these spot diagrams were calculated with the illumination patterns described above. The spots for the NIR camera and for the VIS camera are enclosed in (3-pixel x 3-pixel) sampling boxes.

These spot diagrams must be interpreted carefully because the *vertical* image smear is due to *lateral color*, which is *irrelevant* in cameras that will never be used for direct imaging! Thus, the viewer must compress the spots vertically in his/her mind to assess them in a way that is more nearly appropriate to their intended use as spectroscopic-only cameras.

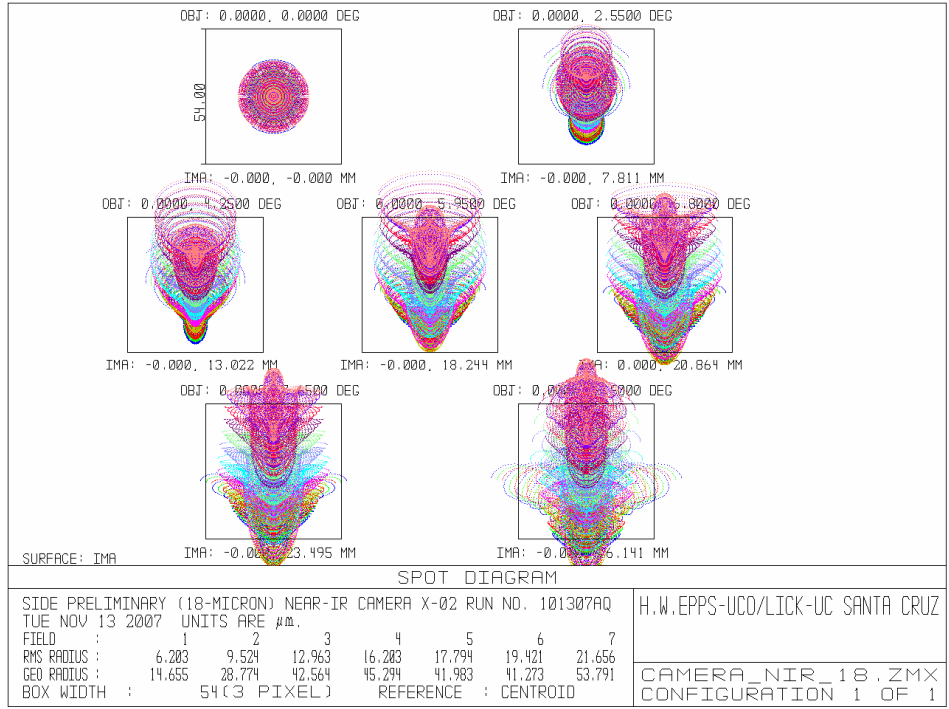

Fig. 7. Spot diagram for the NIR camera (2k×2k @ 18 μm detector).

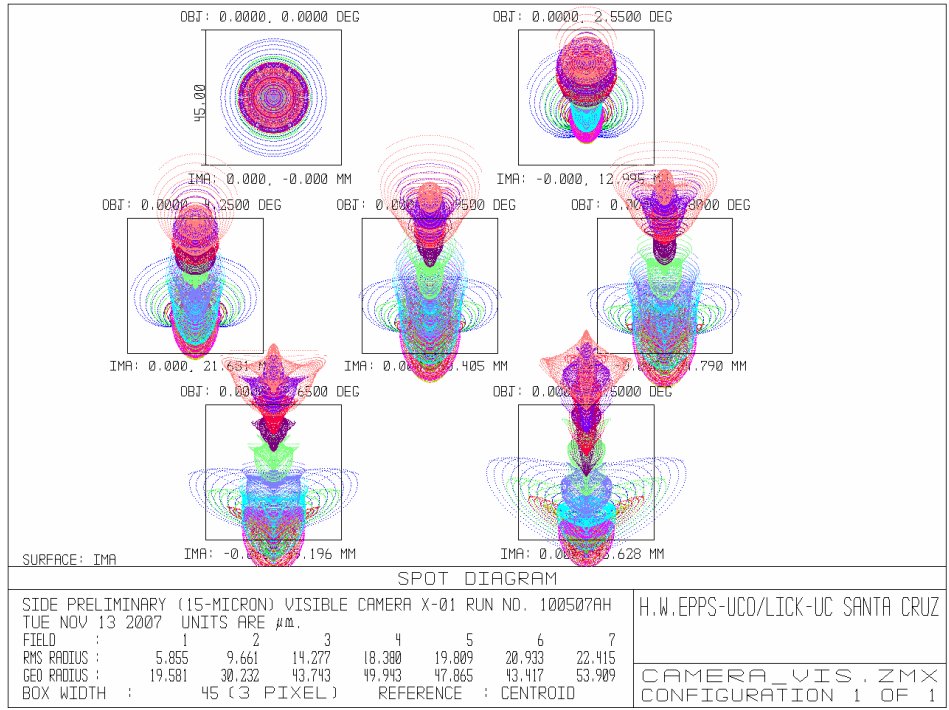

Fig. 8. Spots diagram for the VIS camera.

The subjective point to be made is that indeed, virtually all of the energy will be included within the (3-pixel x 3-pixel) sampling boxes, such that the cameras will certainly be pixel-sampling limited in practice.

## 4. CONCLUSIONS

We conclude that the Dual spectrograph is feasible and requires no challenging or risky technology, and is highly adapted to replicate at a low cost.

## REFERENCES

[1] Epps, H.W. and Elston, R., "Instrument Design and Performance for Optical/Infrared Ground-based Telescopes" Proc. SPIE 4841, 1280-1294 (2002).